\documentclass[12pt,preprint]{aastex}

\begin{document}

\title{Reconnectionless CME eruption: putting the Aly-Sturrock conjecture
to rest}

\author{L. A. Rachmeler\altaffilmark{1},
  C. E. DeForest\altaffilmark{2}, C. C. Kankelborg\altaffilmark{3}}

\affil{\altaffilmark{1}University of Colorado at Boulder, Boulder, CO,
  80304-0391, laurel.rachmeler@colorado.edu; \altaffilmark{2}Southwest
  Research Institute, 1050 Walnut Street, Suite 300, Boulder, CO
  80302; \altaffilmark{3}Department of Physics, Montana State
  University, Bozeman, MT 59717}

\begin{abstract}
We demonstrate that magnetic reconnection is not necessary to initiate
fast CMEs. The Aly-Sturrock conjecture states that the magnetic energy
of a given force free boundary field is maximized when the field is
open. This is problematic for CME initiation because it leaves little
or no magnetic energy to drive the eruption, unless reconnection is
present to allow some of the field to escape without opening. Thus, it
has been thought that reconnection must be present to initiate
CMEs. This theory has not been subject to rigorous numerical testing
because conventional MHD numerical models contain numerical diffusion,
which introduces uncontrolled numerical reconnection. We use a
quasi-Lagrangian simulation technique to run the first controlled
experiments of CME initiation in the complete lack of reconnection. We
find that a flux rope confined by an arcade, when twisted beyond a
critical amount, can escape to an open state, allowing some of the
surrounding arcade to shrink and releasing magnetic energy from the
global field. This mechanism includes a true ideal MHD instability. We
conclude that reconnection is not a necessary trigger for fast CME
eruptions.
\end{abstract}


\section{Introduction}

Coronal mass ejections (CMEs), are large expulsions of magnetic field
and plasma from the solar corona. The kinetic and gravitational potential
energy contained in a CME is around $10^{31-32}$ ergs, making these
events some of the most energetic in our solar system \citep{canfieldenergy,forbesreview,hundhausenetal1994,low1990,low2001}.
It is thought that CMEs derive their energy from the magnetic field
of the solar corona because this field is the only possible source
for such a large reserve of energy \citep[e.g.][]{forbesreview,klimchuckreview,lowreview1996}. 

The flux rope model is one possible pre-eruptive CME configuration. A
flux rope is a length of magnetic field that has been twisted along
its axis, often held in place in the corona by an overlying arcade or
ambient field.  It is thought that cool photospheric plasma can become
be trapped in the center of a flux rope, creating a solar filament or
prominence \citep{vanBallegooijen1989,ridgeway1991,Priest1989}. The
flux rope configuration easily explains the clear three-part structure
seen in many CMEs, specifically those associated with prominence
eruptions \citep{hudsonetal1999}. Because these structures are present
in the low corona where magnetic field is strong and plasma density
is low, they are magnetic-field-dominated. Recently, movies from
\emph{Hinode} have shown interesting dynamics that are not described
by the flux rope model which imply that at least some prominences are
not low-$\beta$ \citep{berger2008}, but the flux rope model remains
useful. The lower coronal environment is also frequently modeled as
being force-free because flow speeds are low, and
$\mathbf{J}\times\mathbf{B}$ is the dominant force in the equation of
motion. Gravity is also frequently ignored because it is a factor of
$\sim5$ weaker than the magnetic forces. Low-$\beta$ flux ropes are
stable when the outward magnetic pressure force is balanced by an
inward-directed tension force. In models, an exterior arcade field is
often added to increase the tension force and keep the flux rope from
simply expanding in length and width as twist is added. The
approximate energy per unit length along the axis, $U$, stored in an
unconfined Gold-Hoyle flux rope \citep{goldhoyle1960} is given
by \begin{equation}
  U=\frac{1}{8\pi^{2}}\frac{\Phi^{2}b^{2}}{ln\left(1+b^{2}R^{2}\right)}\label{eq:flux
    tube energy}\end{equation} where $\Phi$ is the magnetic flux, $R$
is the radius of the tube, and $b$ is the twist parameter such that
$\frac{B_{\phi}}{B_{z}}=br$ \citep{sturrock2001}. As the twist
accumulates, $b$ increases, $\Phi$ is constant and $R$ remains
approximately constant so the total energy increases. When an
unconfined flux rope anchored at both ends on the photosphere
accumulates twist, its equilibrium state is expanded in length
relative to the untwisted state so the twist per unit length does not
necessarily increase with the total twist. Thus, flux ropes that are
confined by an overlying arcade contain more energy because their
length changes very little as $b$ increases.

\begin{figure}
\begin{centering}
  \includegraphics{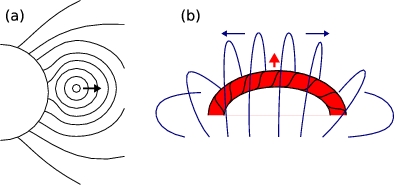}
\par\end{centering}

\caption{Behavior of flux rope expansion in 2.5D (a) and 3D (b).}
\label{fig:2.5d vs 3d}

\end{figure}

The energy stored in the magnetic field is given by the volume
integral of $B^{2}/8\pi$, up to conversion factors. The minimum energy
of a magnetic system with a given photospheric boundary occurs in the
potential, or vacuum field, configuration. As the field is stressed
away from this configuration due to photospheric movements, the energy
is increased above the evolving potential state by an amount commonly
referred to as the magnetic {}`free energy'. If the field reverts to
the potential configuration, this free energy is released and in the
case of solar active regions, is available to drive a CME. If the
reconnection is localized and helicity is conserved, the lowest
accessible energy state may not be potential, so the free energy is an
upper boundary on the amount of energy that can be released.  There is
a global cap on the amount of energy that the magnetic field can
provide. The magnetic virial theorem asserts that the total pressure
force can not exceed the tension force in a stable plasma environment
\citep{priestbook}. The related Aly-Sturrock conjecture states that
the global magnetic energy of a force free field is at a maximum when
the field is completely {}`open'. This refers to magnetic flux that is
anchored at the solar surface and extends radially outward a
significant distance so that, near the Sun, the field appears open
\citep{aly1984,aly1991,sturrock1991}.  Many CME observations show
prominences lifting off of the surface of the Sun, expanding to
several solar radii and leaving behind long radial field lines. If
this conjecture is correct, then the implication is that CMEs which
open large amounts of field must derive the bulk of their kinetic
energy from sources other than the magnetic field because the field
energy is actually greater in the post-CME configuration.  Order of
magnitude analysis has shown, however, than the magnetic field is the
only source of energy that can potentially drive a $10^{32}$erg CME
\citep{forbesreview}. This poses a significant problem to ideal CME
models. In an azimuthally symmetric 2.5-D case, all of the field lines
originally above a prominence-like feature would have to open to
release the filament (Fig. \ref{fig:2.5d vs 3d}a). Thus, in the 2.5-D
case, to have an eruption which results in a net decrease of magnetic
energy, reconnection must be present. There have been studies in which
a 2.5-D field is shown to have magnetic energy exceeding the open
field energy when mass loading is present, but it has not been
demonstrated that these fields can erupt without reconnection
\citep{lowreview1996,fonglowfan,zhanglow}. In three dimensions, the
flux rope is anchored in the photosphere, and the surrounding field
can move away in the direction parallel to the flux rope axis (Fig.
\ref{fig:2.5d vs 3d}b). Reconnection is not necessary in the fully
three-dimensional case, as not all of the field must be open to have
an eruption, only the flux rope opens, and thus the eruption is not
relevant to the hypothesis of Aly and Sturrock because some of the
field remains closed \citep{low1986}.

The full Aly-Sturrock conjecture has yet to be disproved. However, it
is not usually relevant to 3-dimensional models and analysis for the
reasons stated above. A more relevant question which has been asked,
is whether a configuration with \emph{some} open force free field can
contain less energy than a configuration with the same boundary
conditions that is fully closed \citep{low1990}. This question has
been addressed semi-analytically by \citet{woldsonandlow1992} and
\citet{wolfson1993}, who showed that a fully closed field can contain
magnetic free energy above the partially open field threshold, but
they did not demonstrate a release mechanism for this energy. A
demonstration of the ideal evolution of a field whereby free energy is
stored then released, resulting in a partially open state, would once
and for all eliminate the problem posed by the magnetic virial theorem
and the Aly-Sturrock conjecture in a fully 3D case.

\citet{sturrock2001} describe the possibility of driving CMEs with
metastable magnetic fields. They analytically describe one system in
particular as a known metastable state: the previously mentioned
twisted flux rope under an overlying arcade. This system is metastable
because it is stable (due to the confining arcade) against small
perturbations, but the energy of the erupted flux rope is lower than
the contained flux rope. If the rope is tightly wound, it can open and
escape the arcade by herniating through it, leaving the deflated
arcade near the footpoints. The amount of twist needed is not
unreasonable for the solar surface. Analytically, for a flux rope that
is ten times longer than its radius, the rope need only exceed 1.5
total turns about its axis to be in this metastable state
\citep{sturrock1991}.

Numerous simulations exist which model flux rope CME initiation of the
metastable configuration described above
\citep[e.g.][]{aulanier2005,g+f04,titov+demoulin,torok+kliem2004,rousesevetal2003}.
Most of these simulations have found that it is possible to herniate
through the arcade, but they do not agree exactly on how much twist is
needed, or how unstable the resulting configuration is after the onset
of writhe (helical geometry in the central field line) in the flux
rope.  Typically, these codes agree that the the critical twist needed
to erupt is around 1.5 turns, and that it is possible to get an
eruptive event by twisting the footpoints of a flux rope.

CME initiation with these flux ropes has been modeled both with and
without reconnection as the intended primary destabilizing factor.
Theories that do not involve reconnection are referred to as ideal
{}``loss of equilibrium'' models \citep[e.g.][]{rousesevetal2003}.
The initial structure generally undergoes an ideal instability, such
as the MHD kink instability, caused by a large amount of
twist. Another possibility, if mass loading is critical in keeping the
structure contained, is that mass displacement could upset the force
balance and start the CME
\citep{klimchuckreview,fonglowfan,zhanglow}. Other theories, such as
tether cutting \citep{mooreroum} or {}``breakout'' \citep{breakout},
explicitly include reconnection to decrease the strength of the
overlying arcade. In models such as tether cutting and breakout, there
are generally two stages of reconnection. Slow -- Sweet-Parker style
\citep{parker_sweetparker1963} -- reconnection occurs early in the
evolution and destabilizes the system, allowing it to expand. This is
often followed by fast -- Petcheck style \citep{petschek1964} --
reconnection, which releases large amounts of energy in a short time
and is believed to be the primary driver for fast, impulsive CMEs.

Essentially all existing numerical simulations of CME onset use
Eulerian methods, in which a 3-D grid of values is used. With magnetic
fields (and indeed all vector fields and flows), sharp gradients are
not conserved because derivatives are represented as finite
differences.  For magnetic fields, this means that the field will
reconnect if gradients approach the size of the grid whether the
modeler wishes it or not. Techniques such as adaptive mesh refinement
can reduce the rate of numerical reconnection, but can not remove it
altogether. Hence it is not possible to separate the effects of ideal
MHD evolution and magnetic reconnection with an Eulerian grid
code. This can be problematic insofar as the reconnection destabilizes
a metastable system. By switching to a Lagrangian (field-aligned)
formulation, we eliminate all reconnection, allowing study of ideal
MHD instabilities \citep{fluxonpaper}.

With our model, we are able to analyze simplified systems where
topology is locked in and reconnection is not present. Note that we do
not hypothesize that reconnection is not present in the Sun, only that
to have a controlled numerical experiment, the effect of reconnection
must be isolated, and we do this by eliminating it. Our method is
unique in this way, and may offer insights that grid simulations can
not. In particular, we are able to demonstrate the existence of a true
MHD instability that can release free energy into a CME, even with no
triggering reconnection.

\section{Numerical Model}

\subsection{Computational Model}

\begin{figure}
\begin{centering}
\includegraphics[width=3in]{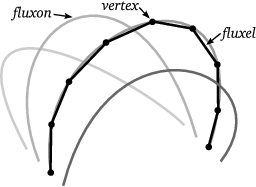}
\par\end{centering}

\caption{Geometry and nomenclature of the FLUX code. Finite-magnetic-flux field
lines are called fluxons, which are broken into linear fluxel segments
joined by vertices.}
\label{fig:fluxels}

\end{figure}

The code used in this work is called FLUX (FieldLine Universal
relaXer) \citep{fluxonpaper}. This quasi-Lagrangian code represents a
three-dimensional field as a collection of \emph{fluxons,} or field
lines with finite magnetic flux. Each fluxon is broken into piecewise
linear segments called \emph{fluxels}, which are joined at vertex
points (Fig. \ref{fig:fluxels}).  To reconnect, a fluxon must be
explicitly broken and connected to another fluxon. With no
reconnection, the code preserves magnetic topology; this is the case
used in the current work. FLUX is coordinate free, so in order to
compare the simulations with the Sun, we assume that our system is
originally the size of an active region, which is a few tens of Mm
across (1 spatial unit = 25 Mm). FLUX is under development and the
version we used for this work is not a full magnetohydrodynamic
code. It does not include the effects of mass or plasma; thus it does
not model dynamics. We are neglecting short time-scale changes of the
system in favor of concentrating on the large scale evolution.

FLUX computes nonlinear force-free equilibrium solutions from a
prescribed initial topology and connectivity by balancing the
components of the Lorentz force, which are resolved as a magnetic
pressure and a magnetic tension force:\begin{equation}
  0=-\nabla\left(\frac{B^{2}}{8\pi}\right)+\frac{\left(\mathbf{B}\cdot\nabla\right)\mathbf{B}}{4\pi}\label{eq:lorentz
    force}\end{equation} The tension force is computed from the
geometry of the other fluxels on the same fluxon (e.g. the angle
between successive fluxels), and the pressure is computed based on the
geometry of the nearest surrounding neighboring fluxels. Each vertex
is moved in the direction of net calculated force until the ratio of
the net force to the sum of the magnitudes of the forces on each
vertex is below a threshold level.  Once all vertices are below this
threshold, e.g. $0.1\%$, the field is deemed to be in equilibrium. (A
more detailed mathematical description is available in
\citet{fluxonpaper}).

Initial conditions in the code consist of a planar line-tied
photosphere-like boundary with a set connectivity. The footpoint of
each fluxon can be moved independently to simulate photospheric
motions. After each footpoint movement, the field is allowed to relax
to equilibrium before the next movement occurs. In this way, it is
possible to create a quasi-static evolution of equilibrium states. The
simulation is bounded at the top by an open hemisphere. Fluxons that
intersect this surface are free to move around on it. Closed loops
that approach the surface are truncated and become two separate
fluxons that then move independently. Open fluxons move to equalize
magnetic pressure (there is no curvature on the final fluxel), which
has the effect that the magnetic field at the upper boundary is normal
to the surface.

\subsection{Simulation set up}

\begin{figure}
\begin{centering}
\includegraphics[width=2.75in]{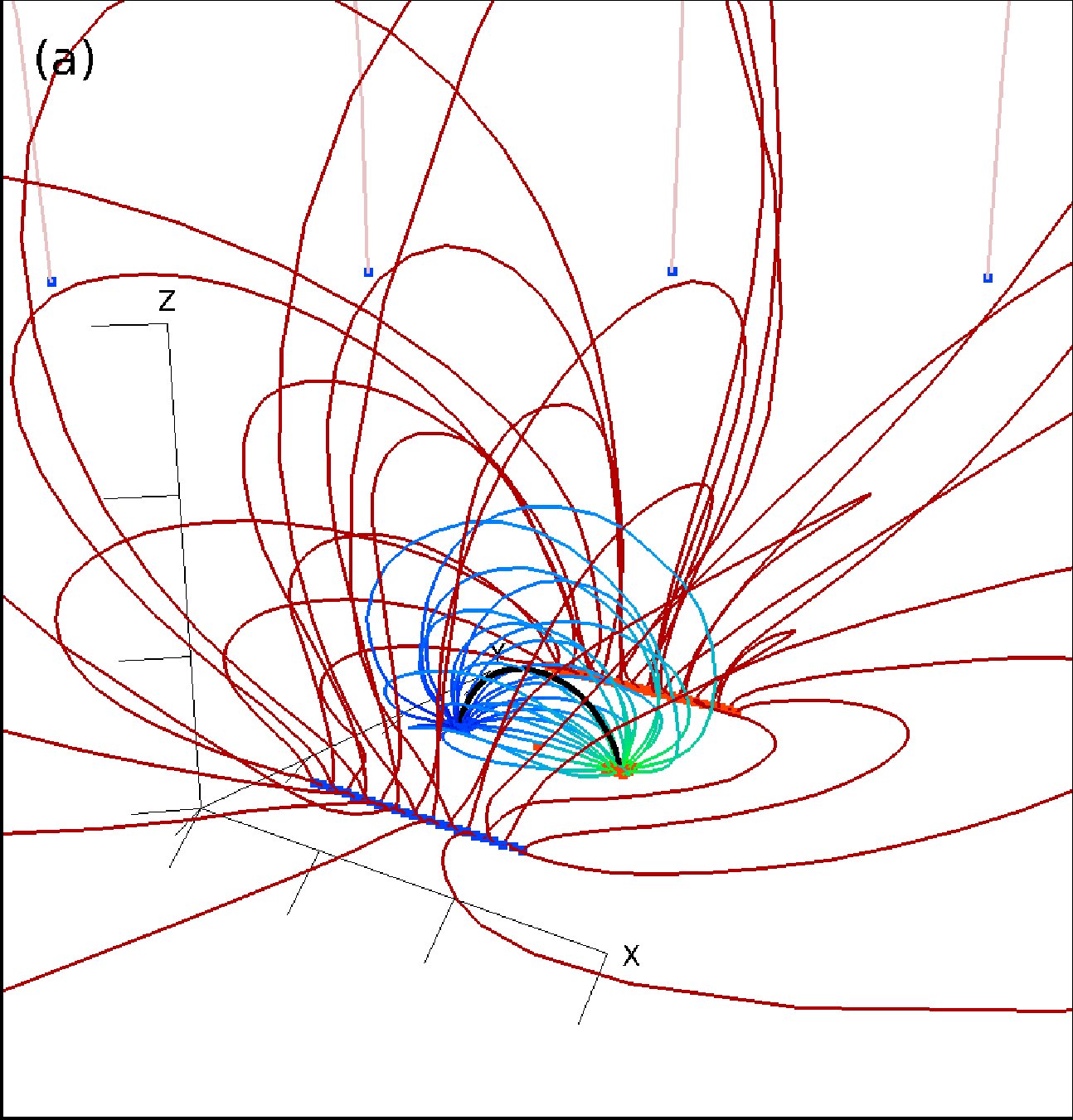}\includegraphics[width=2.75in]{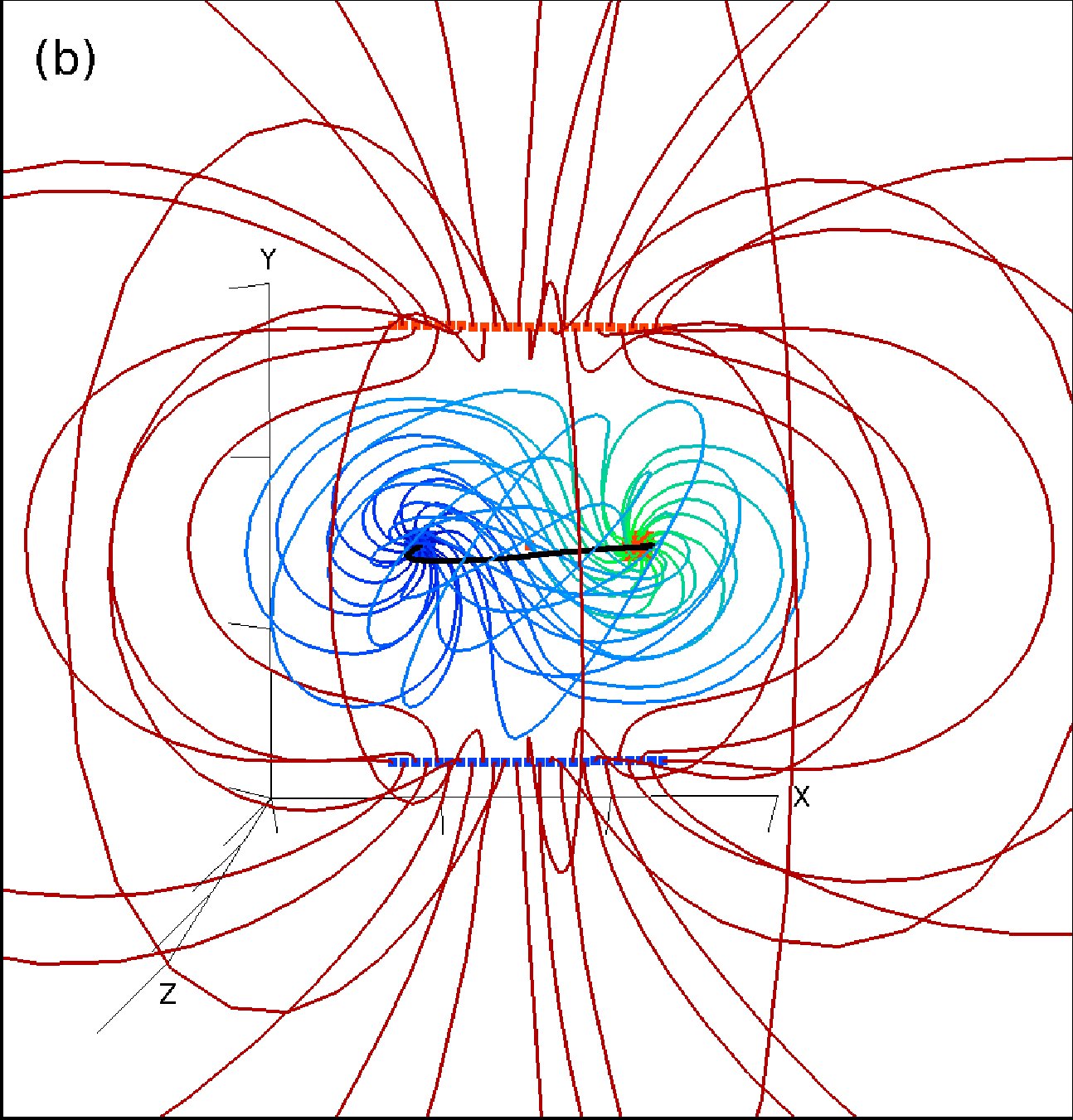}
\par\end{centering}

\begin{centering}
\includegraphics[width=2.75in]{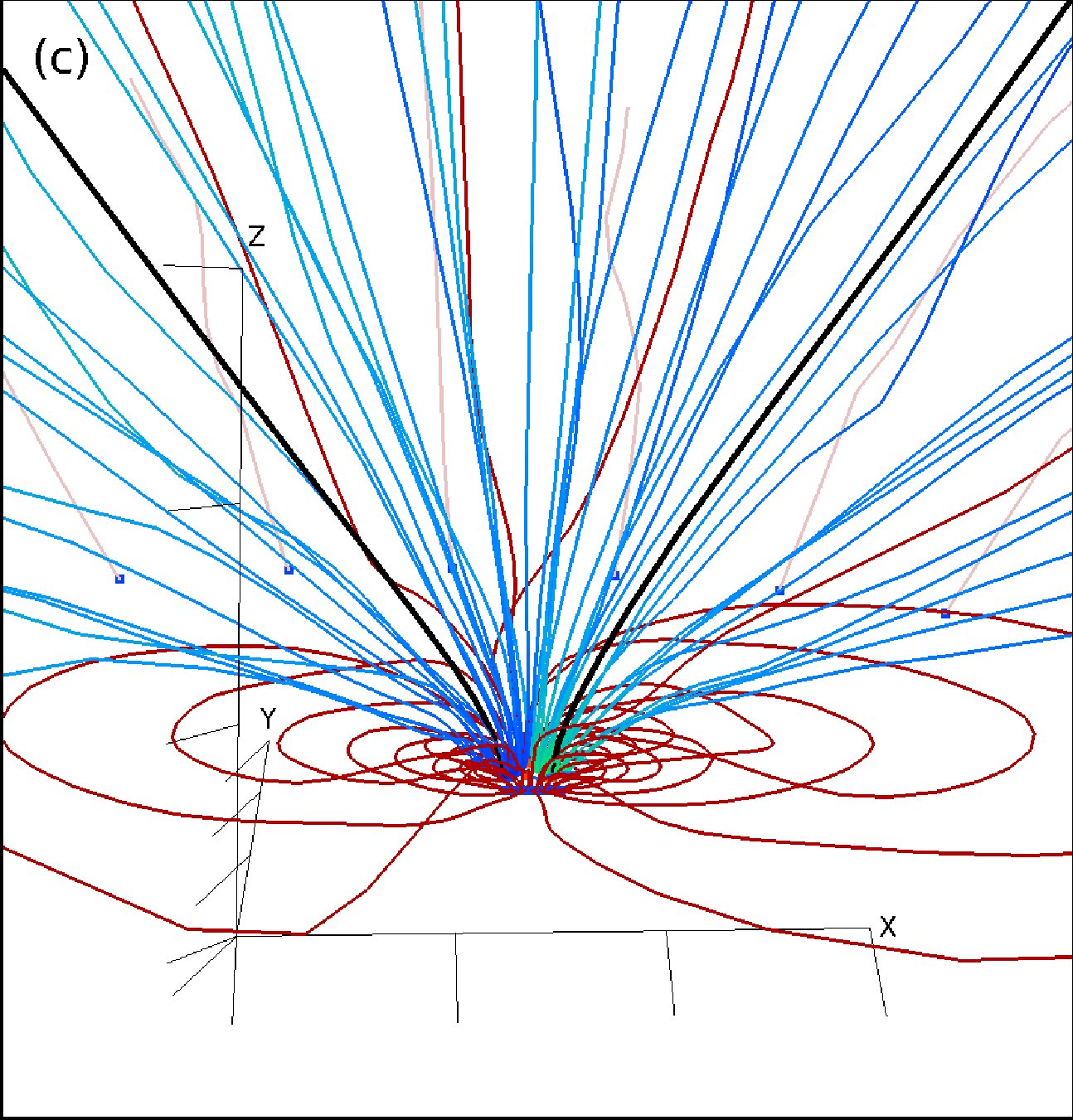}\includegraphics[width=2.75in]{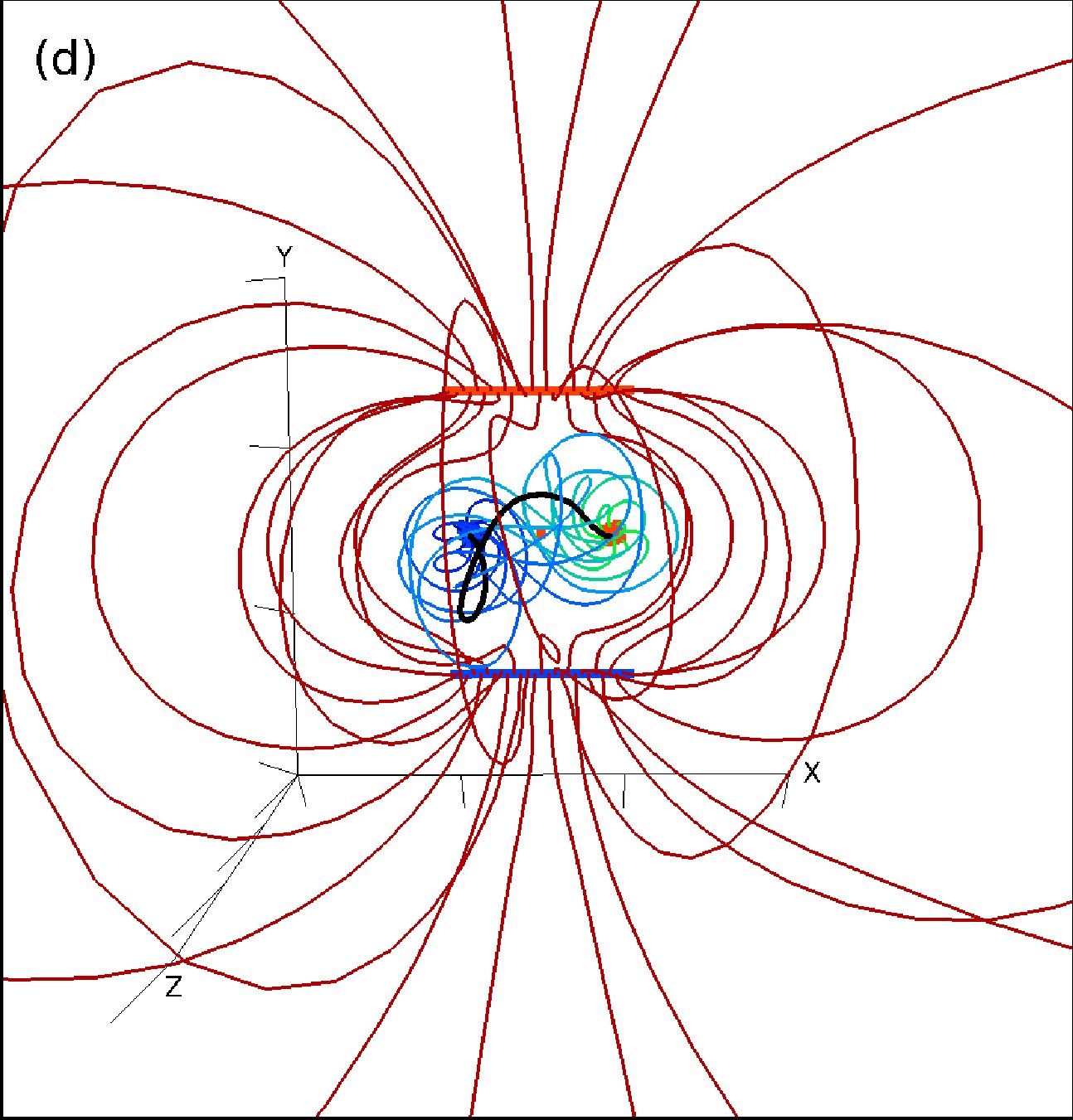}
\par\end{centering}

\caption{\label{fig:sequence} (a)-(c) Initial simulation set-up and
  evolution of a system with 75 fluxons where the arcade is 25 fluxons
  shown in red and the flux rope is 25 fluxons in shown in a gradient
  from blue to green to show directionality. The central fluxon is in
  black to highlight the flux rope axis. The outer ring of 25 fluxons
  is shown in light red. (a) The equilibrium system with no twist, (b)
  with 1 turn, which has not kinked, (c) with 1.43 turns which has
  herniated and is open at the upper boundary. (d) Shows a different
  system with 9 fluxons in the flux rope and 3 units of flux (27
  fluxons) in the arcade at 2.1 turns, which demonstrates kinking
  without herniation. Note that the systems with one unit of arcade
  flux do exhibit a slight kink while confined as evidenced by a small
  distortion in the central fluxon, but the systems with higher arcade
  flux exhibit much stronger kink before herniation.}

\end{figure}

The simulated systems consist of a flux rope, an overlying arcade, and
an outer ring of open field lines. Figure \ref{fig:sequence}(a) shows
this set-up. The fluxons are tied to a planar lower boundary and
evolve with a prescribed surface motion. The central flux rope is
twisted incrementally in a solid body rotation pattern by four degrees
each step and allowed to relax to equilibrium.The flux rope footpoints
are set at 2 spatial units apart, or 50 Mm.

One difficulty in examining these results has been with the energy
calculation. Our code calculates the energy of every fluxel based on
the cross sectional area it occupies and its length. In regions where
the fluxons are close to each other, this method works extremely well,
but it has more trouble for the outer-most fluxons in a system.  We
call this the {}`last-fluxon' problem and it is discussed by
\citet{fluxonpaper}. The volume that the last fluxon occupies is
infinite, so it cannot be treated as small, violating the
approximation used by the code. Once the system has herniated and
expanded fully, the number of last-fluxons is much greater,
exaggerating the difficulty in determining the post-eruption
energy. The open hemispherical surface at $35\, R_{\odot}$ and an
outer ring of fluxons were added to alleviate this problem. As a
consequence of the surface, the field opens once it encounters the
hemisphere and expands to fill the volume.

The footpoints of the outer ring remain stationary throughout the
simulation. This outer ring is present to minimize the {}`last fluxon
problem' because with the ring, none of the arcade or flux rope
fluxons will be a last fluxon. The ring is positioned far from the
rope and the arcade, about $2\, R_{\odot}$ away, so it does not effect
the evolution. The outer boundary at a radius of $35\, R_{\odot}$ is
much farther than the physical regime of applicability of FLUX. The
transition to solar wind occurs at $\sim4\,
R_{\odot}$\citep{parker1960,kohl1997}, so at most, our results have
physical meaning up to this height.

We performed several simulations with varying numbers of fluxons in
the flux rope and the arcade, each with the same basic set-up. The
flux rope consists of 9, 16, 25, 36 or 49 fluxons arranged in a square
on the photosphere. The footprint of the flux rope is the same size
in each case. For the rest of this paper, a \emph{unit} of magnetic
flux refers to the amount of magnetic flux associated with the flux
rope, or the number of fluxons in the flux rope. In these simulations,
the flux rope and the ring consist of one unit of flux, and the arcade
contains one, two, or three units.

\section{Simulation Results}

\begin{figure}
\begin{centering}
\includegraphics[scale=0.5]{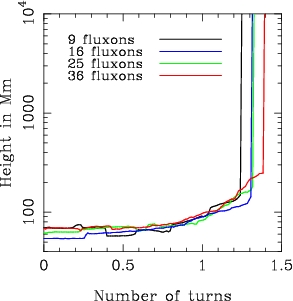}
\par\end{centering}

\caption{\label{fig:height plot} The flux rope remains well confined
  as it is twisted until it herniates rapidly through the arcade. The
  data shown are for systems with one unit of flux in the arcade.}

\end{figure}

In all cases, we find that the flux rope herniates through the arcade
after a certain amount of twist has been applied, entraining a few
arcade fluxons with it as it goes. Figure \ref{fig:sequence}(a)-(c)
demonstrates a typical sequence of events for a case of a 25 fluxon
flux rope and a one-unit magnetic flux arcade. First, the flux rope
twists about its central axis under the arcade. After about 1.4 turns
have been applied to the flux rope (for one unit of arcade flux), the
rope herniates. In this case, the flux rope does not significantly
kink -- the central axis remains mostly un-twisted -- but in the case of
a stronger arcade, the flux rope does kink before it herniates.  The
stronger the arcade, the flatter the flux rope is, and the more twist
is required to initiate herniation. Figure \ref{fig:sequence}(d) shows
the three-unit arcade system after it has undergone writhe; the black
central fluxon is no longer straight. The onset of kink does not
trigger the herniation through the arcade. The flux rope continues to
twist and writhe until it begins to herniate. In every case, after the
onset of herniation, the flux rope expands rapidly while the arcade
deflates. Figure \ref{fig:height plot} shows a plot of height of the
flux rope vs. twist imparted, for various fluxon densities with one
unit of arcade flux. The expansion occurs extremely rapidly, within
one equilibrium step, once the flux rope breaks through the arcade.

We label this rapid expansion as an {}``eruption,'' because the size
of the flux rope increases by a factor of 100 or more and breaks
through the open surface in a single equilibrium time step, while
releasing a non-trivial amount of energy. Thus, these large expansions
are deemed eruptions and the flux rope fluxons are labeled as open.

The amount of twist needed to herniate through a given arcade strength
varies with the number of fluxons used to represent the field
(Fig. \ref{fig:height plot}; this may be indicative of grid effects
that are setting an unstable twist level or seeding the
instability. Because of the discrete nature of fluxons, the system is
not always symmetric, and this probably accounts for some uncertainly
in the critical twist. There is not always a consistent trend with
fluxon density, and so there may be other reasons behind this
behavior.

\begin{figure}
\begin{centering}
\includegraphics[width=2.25in]{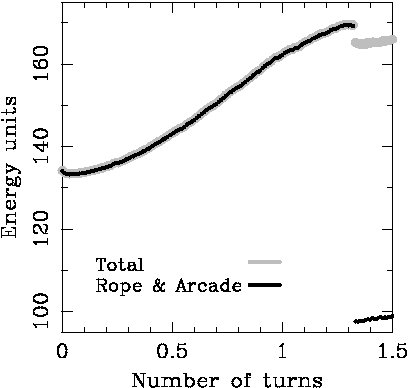}
\par\end{centering}

\caption{\label{fig:energy} A typical energy plot for the simulations
  run.  This one shows the energy evolution of a 25 fluxon flux rope
  with a one unit flux arcade. As the flux rope is twisted under the
  arcade, the energy increases steadily. Once the system herniates,
  the energy loss is substantial. Note that the energy in the flux
  rope and arcade only (black line) drops to lower than the initial
  level. This is because most of the energy in the system at that
  point is in the outer ring fluxons which are highly stressed toward
  the horizontal.}

\end{figure}

The energy of the final erupted state is less than the energy of the
confined flux rope (Figure \ref{fig:energy}). Note that the presence
of the open boundary does not skew these results. The magnetic energy
that escapes through the boundary would not be available to drive the
CME in any case because it is present beyond the transition to the
solar wind. In the latter stages of the expansion, the twist per unit
length in the flux rope is small, and hence the free energy is low. It
is the initial expansion that drives the CME, not the later
expansion. Compared to the physical case, we overestimate the final
magnetic energy of the system because our boundary is much farther out
than the transition. A significant amount of energy is available to
drive a CME, even without reconnection.

This simulation allows us to put a strong upper and lower bound on
the amount of energy that is released with the eruption. Of the free
energy injected, $12.5\%$ is lost after herniation. The energy calculated
in the ring field after herniation is an overestimate, and the energy
in the flux rope and the arcade is an underestimate for the following
reasons. The ring field was added so that all of the last fluxons
were that ring. As stated earlier, we do not trust the energy calculation
for the ring field, especially considering that before herniation
less than $1\%$ of the system's energy was in the ring field compared
to $\sim40\%$ after herniation. Also, on a spherical solar surface,
the ring field would be farther away than the disk limb if the flux
rope were at disk-center, and consequently would not be highly sheared
away from radial after the eruption, so it would not store much more
energy than it had initially.

Because of the unreliability of the final energy in the ring field, we
also looked at the energy in only the flux rope and the arcade. In
this partial system, the final energy is less than the initial
potential energy in part because some of the energy is carried through
the open boundary and lost from the calculation, and in part because
this limited system does not account for any background solar field
that may be deformed by the CME.

Despite these effects, we are able to conclude that a significant
amount of energy is released and could be used to drive an impulsive
CME. The best way to resolve the energy would be to run a similar
system as a full-Sun simulation in spherical geometry so that there
are no last-fluxons, and include an estimate of the energy outside the
upper boundary with a force free field extrapolation. This future work
may be able to determine quantitatively how much magnetic energy
available; a figure which is highly dependent on the geometry of each
event.

\section{Discussion}

Our simulations show that reconnection is not necessary to initiate a
CME and that impulsive CMEs may be possible without explosive
reconnection.  This theory is not new; it was originally published by
\citet{sturrock2001}, who describe the existence of metastable states,
specifically a system similar to the one we have studied. Since then,
other solar physicists have studied this system computationally
\citep[etc.]{g+f04,torok+kliem2004,aulanier2005}.  The results from
these studies show that a highly twisted flux rope can herniate
through a confining magnetic arcade and reconnect into a plasmoid,
causing an eruption. However, this is the first study of this system
in the complete lack of reconnection. In general, our results agree
with those of other research groups.

The fact that many of these simulations, including ours, agree that
about 1.5 total turns is needed to herniate through an arcade implies
that reconnection is not greatly important to the overall stability of
the system. If it were, then we would expect our ideal simulation to
support significantly more twist and therefore release more energy
than the dissipative simulations. The exact value of the critical
twist will depend on the configurations of the system: strength of the
arcade, width of the flux rope, the twist profile within the flux
rope, etc. But even with these variables, we conclude that highly
twisted flux ropes can not easily be confined by external field, even
when the reconnection rate is extremely small.

Current research on the twist available photospheric fields indicates
that there may be an excess of one full turn available in many active
regions \citep{leka}. This implies that many pre-eruptive active
regions may be on the brink of an ideal instability when they flare or
erupt, regardless of the eventual trigger mechanism.

Our results together with the results of \citet{woldsonandlow1992} and
\citet{wolfson1993} can finally put to rest the concerns that the
Aly-Sturrock conjecture have created over the initiation of
CMEs. Although it has not been proven that a closed force free field
can have more energy than a fully open one, that question is not
relevant in the complex three dimensional system that is our Sun. A
more appropriate question is whether a closed force free field can
contain more energy than a configuration with some open field given
the same lower boundary and connectivity. \citet{woldsonandlow1992}
began to answer that question with semi-analytic techniques and
successfully showed that it was possible. We have proven that it is
possible to transition between closed and partially open field while
still releasing free energy without reconnection and without the need
for gravitational or other non-magnetic confinement. At the beginning
of the century there was, {}``still no model which demonstrates that a
partly open magnetic field can be achieved solely by a loss of ideal
MHD equilibrium or stability.''  \citep{forbesreview}. Happily, this
statement is no longer true.

\acknowledgements
This work was funded by NASA's LWS-TR\&T
  program. FLUX is open source software available from
  http://flux.boulder.swri.edu. Thanks to the PDL development team
  http://pdl.perl.org. We also owe thanks to Spiro Antiochos, Bernhard
  Kliem and Zoran Mikic for valuable discussions of the fluxon
  technique.

\bibliographystyle{plainnat}

\end{document}